\documentclass[journal=jacsat,manuscript=article]{achemso}

\usepackage[version=3]{mhchem} 

\author{Armin Abdehkahka}

\author{Craig Snoeyink}

\email{craigsno@buffalo.edu}
\phone{+1 (716) 645-1468}

\affiliation[University at Buffalo]
{Department of Mechanical and Aerospace Engineering, University at Buffalo}

\title[Localization of Ultra-dense Emitters with Neural Networks]{Localization of Ultra-dense Emitters with Neural Networks}

\abbreviations{IR,NMR,UV}

\begin{document}







\let\elskeyword\keyword

\begin{abstract}
  Single-Molecule Localization Microscopy (SMLM) has expanded our ability to visualize sub-cellular structures but is limited in its temporal resolution. Increasing emitter density will improve temporal resolution, but current analysis algorithms struggle as emitter images significantly overlap. Here we present a deep convolutional neural network called LUENN which utilizes a unique architecture that rejects the isolated emitter assumption; it can smoothly accommodate emitters that range from completely isolated to co-located. This architecture, alongside an accurate estimator of location uncertainty, extends the range of usable emitter densities by a factor of 6 to over $\mathrm{31 ~ emitters/\mu m^{2}} $ with reduced penalty to localization precision and improved temporal resolution. Apart from providing uncertainty estimation, the algorithm improves usability in laboratories by reducing imaging times and easing requirements for successful experiments.
\end{abstract}

\section{Introduction}
Super-resolution methods such as STORM \citep{storm}, PALM \citep{palm}, STED \citep{sted}, and others \citep{gustafsson_surpassing_2000,hell_properties_1992,sharonov_wide-field_2006} have revolutionized biological science by visualizing biological processes beyond the diffraction limit \citep{liu_super-resolution_2022,lelek_single-molecule_2021}. STORM and PALM, both of which are based on localization microscopy have, in particular, made an impact due to their flexibility and high resolution at the cost of poor temporal resolution and computationally intensive data analysis \citep{sage_super-resolution_2019}. Algorithms based on deep learning and Convolutional Neural Networks (CNNs) \citep{liu2018fast,boyd2018deeploco,nehme2020deepstorm3d,decode} have made great strides towards reducing both of these downsides, decreasing analysis times by orders of magnitude and providing gains in temporal resolution through increasing the density of emitters capable of being analyzed in each image.

Improving temporal resolution by increasing acceptable emitter density for AI algorithms has come at a cost to precision and accuracy, with precision rapidly deteriorating alongside emitter density \citep{liu2018fast,boyd2018deeploco,nehme2020deepstorm3d,decode}. This effectively creates a limit to the usable emitter density, beyond which point the resulting localization precision and detection accuracy are no longer acceptable. We hypothesize that this limit is a result of an isolated emitter assumption built into the architecture of current state-of-the-art CNN localization algorithms; they are constructed and trained to return discrete locations corresponding to the 3D locations of individual emitters. As emitter density increases and it becomes increasingly likely that emitter images overlap significantly, these neural networks struggle to return coherent results. 

Convolutional Neural Network (CNN) based localization algorithms, while computationally efficient, also uniquely suffer from a bias at low signal to noise ratios that coerces locations towards the centers of pixels \citep{liu2018fast,boyd2018deeploco,nehme2020deepstorm3d}. This so-called checkerboard effects \citep{artifacts,checkerboard} results in noticeable checkerboard-like pattern in reconstructions that follow the grid of pixels used to acquire the image. There has been some success at mitigating its impact through filtering, though at the expense of recognition accuracy \citep{decode}. We hypothesize that these artifacts are a result of using certain types of pooling and upsampling operations \citep{Fixed} and can be resolved by the appropriate choice of upsampling method \citep{interpolation}, combined with sub-pixel convolution \citep{subconv}.

We present a novel architecture for the Localization of Ultra-high density Emitters using Neural Networks (LUENN), which is designed to make no assumptions about the isolation of emitters and one in which no bias exists at low signal to noise ratios. This is accomplished by lowering the burden placed on the neural network; instead of requiring it to return discrete emitter locations, the output of this neural network is a pair of upsampled grids which contain smoothly varying functions representing the likelihood of an emitter's lateral location and its depth. This output design, alongside avoiding nearest neighbor interpolation when upsampling between layers, results in a neural networked based localization algorithm that can analyze dramatically higher emitter densities at minimal costs to precision, with no bias at low signal to noise ratios (SNRs), and which can provide accurate estimates of emitter localization uncertainty. 

\section{Method}
\begin{figure}[!h]
\centering
\includegraphics[width=0.95\textwidth]{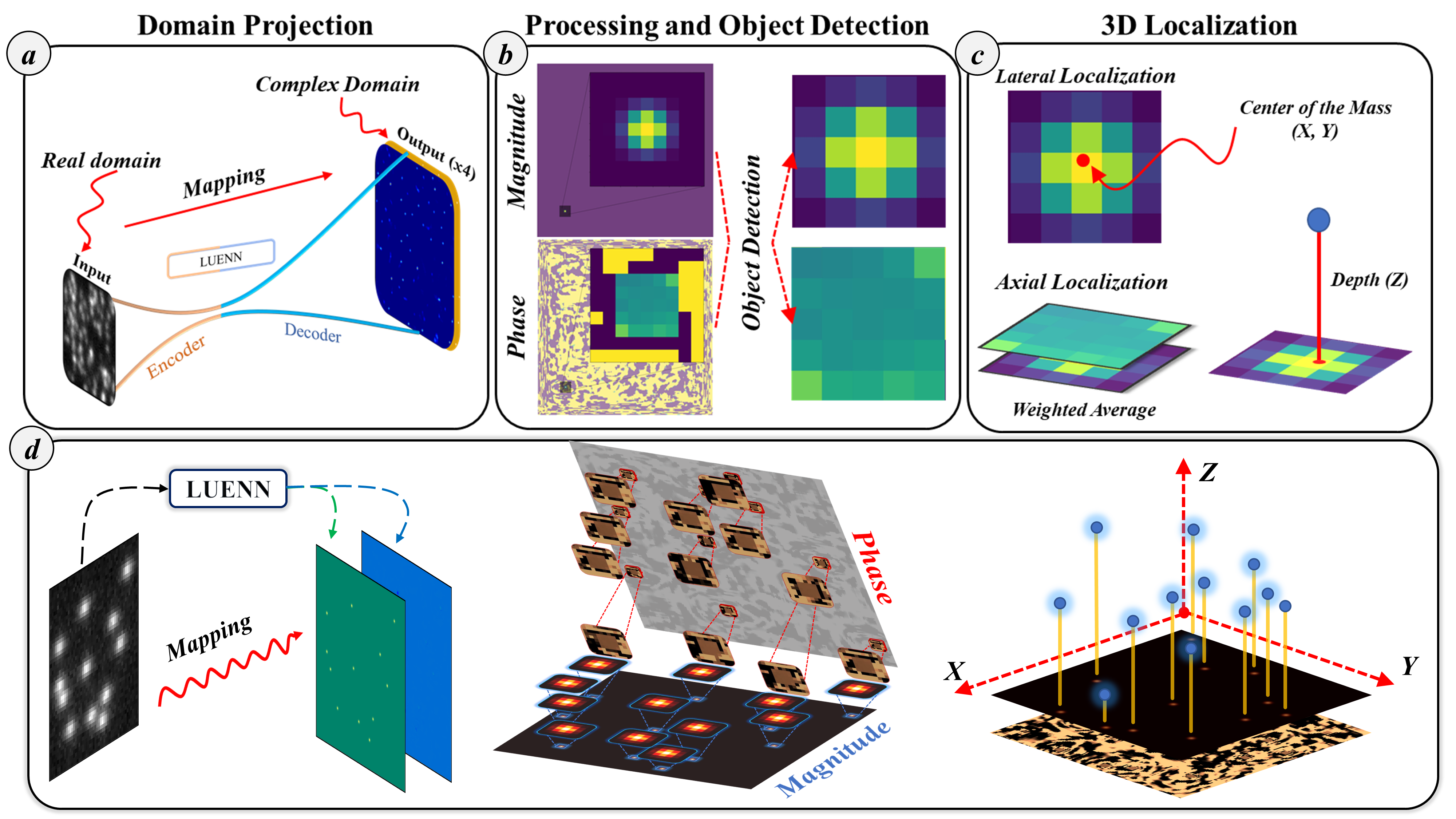}
\caption{\textbf{Image processing workflow by LUENN.} \textbf{a.} Image-to-image transformation and real-to-complex domain projection. \textbf{b.} Object detection and inverse transformation through phase and magnitude calculations. \textbf{c.} Lateral and axial localization of the detected objects. \textbf{d.} A full-processing steps for a low-resolution simulated frame.}\label{fig:work-flow}
\end{figure}

LUENN is an auto-encoder model that incorporates a new output for the simultaneous detection and localization of the emitters in the frame. It is designed based on the U-Net architecture, which is utilized in numerous localization algorithms \citep{Unet,mockl_accurate_2020,hyun_development_2022,zelger_three-dimensional_2018-1} and consists of an encoder and decoder stages as depicted in Figure \ref{fig:work-flow}. The decoder stage integrates the encoded spatial information and maps the output onto two super-resolution ($\times4$) images in the complex domain as shown in Figure \ref{fig:work-flow}a. Lateral and depth information for emitters is coupled through the complex representation layers to facilitate training, paralleling the coupling of lateral and depth localization information through depth-dependent changes to the point spread function necessary for 3D localization. Similar to the complex analysis for the digital signal processing \citep{digital}, the norm of these two channels includes 2D Gaussian distribution for each emitter in the sub-pixel scale with its center-of-the-mass set as the 2D location (in X-Y coordinate) and the corresponding phase represents the depth of the emitter in the axial direction (z). See supplementary Figure 1 for more details of LUENN architecture.\\

This representation frees the neural network from estimating discrete emitter locations and instead asks it to estimate an analog of the likelihood of an emitter and their depth at each sub-pixel location.  Training of LUENN utilizes a supervised learning approach based on the realistic frame simulation method presented by Spieser et al. \citep{decode}. In the training, emitter ground truths distributions are summed allowing for smooth transitions from completely separated to perfectly co-located emitters. 

Introducing this flexibility into the neural network output necessitates additional post-processing to estimate emitter locations. Depicted in Figure \ref{fig:work-flow}c, this takes the form of peak finding and sub-pixel localization in the upsampled lateral likelihood image. Sub-pixel interpolation of the peak is used to estimate the lateral location of the emitter \citep{li1999modified}. The depth of the emitter is then estimated from the phase image, averaged and weighted by the corresponding lateral likelihood peak intensity. 

Another advantage of LUENN's output design is that it reveals the relative certainty and prediction confidence level of the neural network in its results; output distributions for emitters with high uncertainty tend to be more broad amongst other characteristics. We trained an additional convolutional neural network to analyze the output of the primary neural network and provide estimates of the x, y, an z mean population error. The result is an accurate estimate of the underlying  location population statistics that can be used to efficiently filter results to optimize the trade-off between precision and localization accuracy for a given experiment. 

\begin{figure}[!h]
\centering
\includegraphics[width=.6\textwidth]{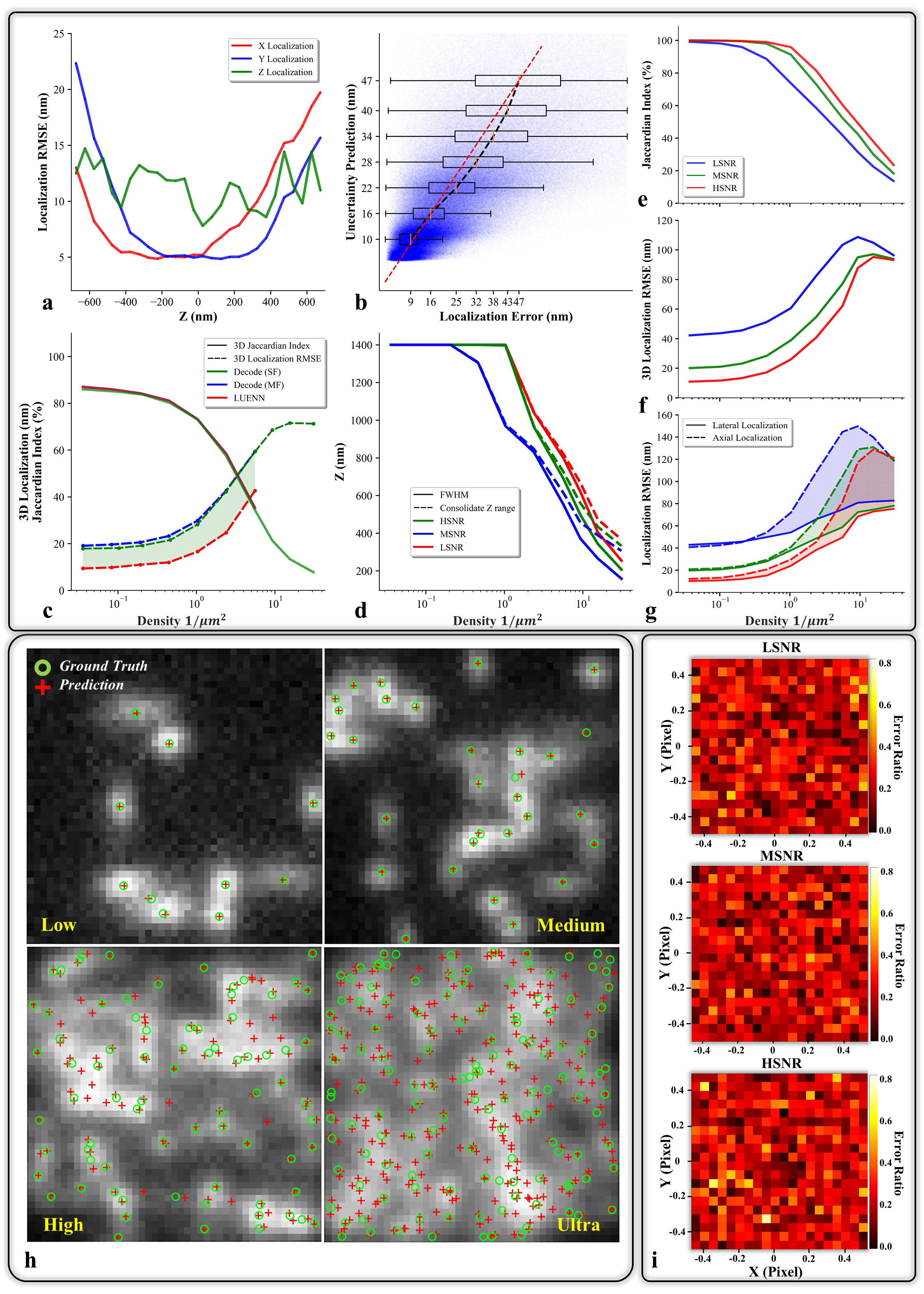}
\caption{\textbf{LUENN Results.} \textbf{a.} Single Emitter Localization Results with high SNR. \textbf{b.} Demonstration of ability to estimate mean error of underlying emitter population. The results are for simulated frames with relatively high density, 0.46 to 1.03 $\mathrm{emitter/\mu m^2}$ and random background intensity levels. The error bars represent the standard deviation of the localization errors of the underlying emitter population. \textbf{c.} Quantitative comparison of the LUENN's detection accuracy (JI) and localization precision (RMSE) with DECODE single frame (SF) and Multi-frame (MF) for medium SNR. The comparison was carried out using variation of simulation parameters (see Figure \ref{fig:decode_compare}). \textbf{d.} Usable measurement depth for astigmatism imaging modality as emitter density varies. \textbf{e-g.} Increased detail on LUENN localization performance with emitter density, h) visualization of LUENN localization results for a range of emitter densities up to $\mathrm{31~emitter/\mu m^2}$. \textbf{h.} LUENN detection and localization performance in example simulated frames with low, medium, high, and ultra-high densities. (i) demonstration of lack of pixel level localization bias in low, medium, and high signal to noise ratio localizations.}\label{fig:results}
\end{figure}

The result of these architecture design choices is an algorithm that can produce precise localization estimates even in ultra-high dense emitter images. Figure \ref{fig:results} shows several measures of performance of LUENN demonstrating, for example, that absent of filtering it can maintain high detection accuracy with minimal penalty even at emitter densities of $\mathrm{1~emitter/\mu m^2}$ in Figure \ref{fig:results}e. Beyond $\mathrm{1~emitter/\mu m^2}$ the detection accuracy, here measured using the Jacardian Index (JI) \citep{sage_super-resolution_2019}, steadily falls off for low, medium, and high signal to noise ratios as expected while the localization precision exhibits novel behavior for a localization algorithm. As emitter density increases, the localization precision, here measured using root-mean-squared-error (RMSE) \citep{sage_super-resolution_2019}, increases and then levels off. To our knowledge, all other localization algorithms maintain a monotonic increase in localization precision as emitter density increases, effectively producing a barrier beyond which increased emitter density is not practical. For LUENN, this barrier still exists but is driven by detection accuracy of emitters and is nearly five times higher than current state of the art AI driven localization algorithms like DECODE, i.e. around $\mathrm{31~emitter/\mu m^2}$.

The robustness of this algorithm in the face of increasing emitter density results in improved temporal resolution with minimal penalty to localization precision. For example, in a low signal to noise ratio experiment with emitters randomly distributed within the frame if emitter density is increased from $\mathrm{1~emitter/\mu m^2}$ to $\mathrm{31~emitter/\mu m^2}$, over three times more emitters will be positively localized and with 3D localization precision increasing from 51 nm to 79 nm. This suggests that three times fewer frames need to be acquired to produce a super-resolution reconstruction. 

While this potential three fold improvement in temporal resolution is welcome, it is important to note that the flexibility afforded to users is enormous. Both LUENN localization precision and temporal resolution are relatively insensitive to emitter density at high and ultra-high densities. Doubling emitter density from $\mathrm{15~emitter/\mu m^2}$ to $\mathrm{30~emitter/\mu m^2}$ will result in roughly the same number of positively localized emitters and no penalty in localization precision. As a result, emitter densities between $\mathrm{15~emitter/\mu m^2}$ and $\mathrm{30~emitter/\mu m^2}$ result in very similar experimental outcomes. 

This improvement in ultra-high density emitter localization precision comes without bias in localization within pixels as can be seen in Figure \ref{fig:results}d. Neural-network based localization algorithms that utilize upsampling within their architecture have traditionally suffered from pixel-level biases in low signal to noise ratio experiments\citep{decode}. State of the art AI based methods like DECODE have successfully removed these errors by increasing filtering, which necessarily reduces their detection accuracy. Using bi-linear interpolation in upsampling layers instead of nearest-neighbor within the CNN removes this bias and the necessity for aggressive filtering. Ultimately, this improves both accuracy and precision of the measurements and minimizes user decisions in filtering. 

\begin{figure}[!h]
\centering
\includegraphics[width=.9\textwidth]{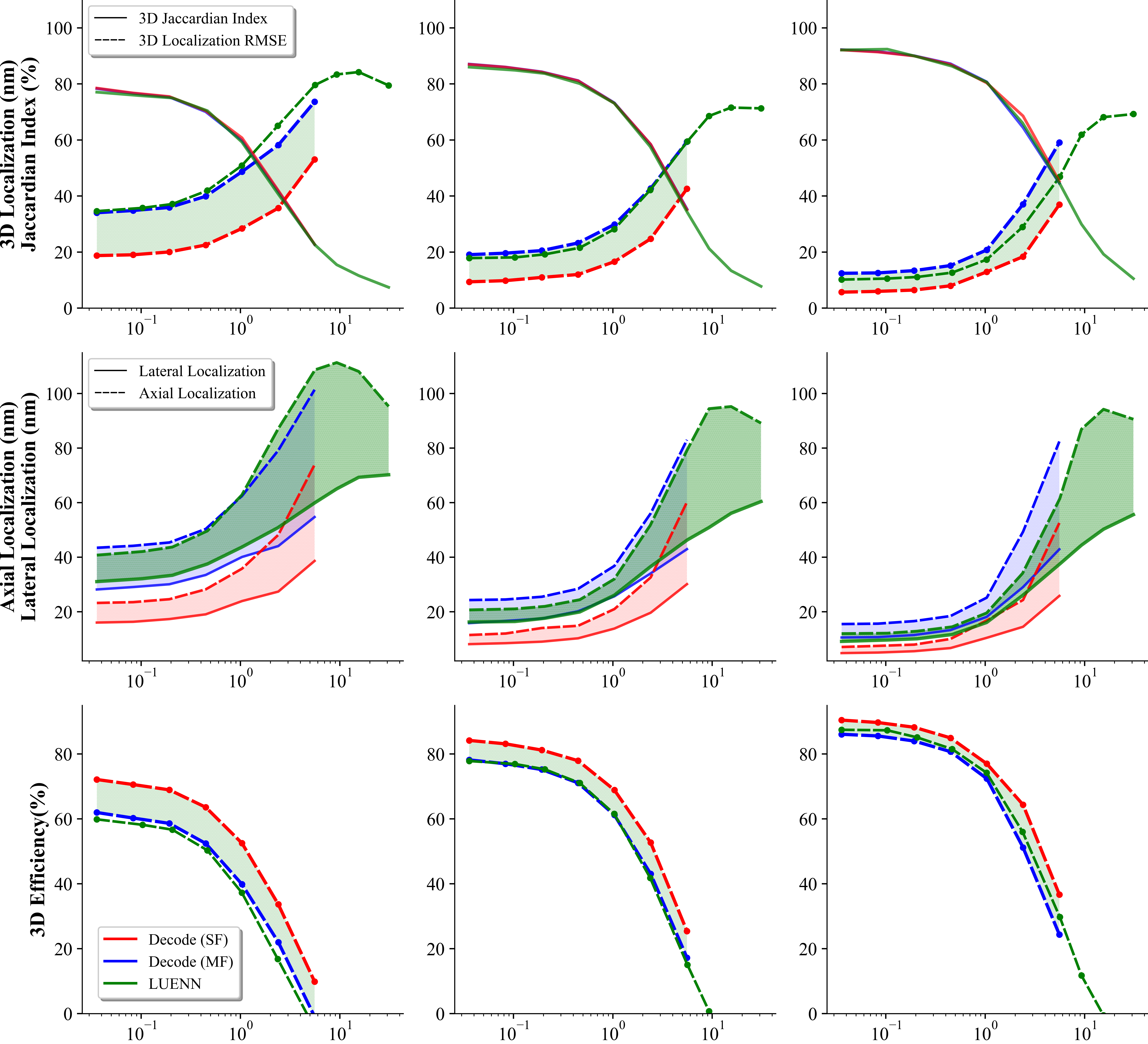}
\caption{Quantitative comparison of LUENN performance with DECODE, both single and multi-frame, at three different SNRs of low, medium, and high (left, center, and right respectively). In each mode, we matched the 3D JI of LUENN, solid lines at the first row plots, to published DECODE results to demonstrate the unique behavior of 3D localization RMSE which begins to plateau at ultra-high emitter densities and matches or exceeds the performance of multi-frame DECODE at emitter densities at or above $\mathrm{5~emitter/\mu m^2}$. The last row shows that the 3D efficiency of LUENN lies above the single frame and below the multi-frame DECODE results except at ultra-high emitter densities where it begins to match and then exceed it.}\label{fig:decode_compare}
\end{figure}

Filtering decisions are made straightforward due to the accurate estimator of localization uncertainty. Results from the uncertainty estimator are shown in Figure \ref{fig:results}b as a scatter plot of the estimated mean error of each localization as a function of the actual error. The mean of the actual errors closely tracks the estimated  mean localization error and successfully estimates the underlying population statistics of each individual emitter based on features from the LUENN network output. This uncertainty estimate can then be used by users to efficiently filter localization results, trading improved uncertainty for detection accuracy (JI) if necessary and is used throughout this work. For example, in the following comparison with DECODE results, the uncertainty threshold was varied to match the detection accuracy in Figure \ref{fig:decode_compare} with those of published DECODE results \citep{decode}.

These results compare favorably with the current state-of-the-art, AI enabled DECODE algorithm as shown in Figure \ref{fig:decode_compare}. LUENN has slightly worse performance at low SNR and better at high SNR at emitter densities less than $\mathrm{5~emitter/\mu m^2}$. Above $\mathrm{5~emitter/\mu m^2}$ there are no published results to compare to, but it is notable that LUENN breaks the trend of increasingly poor precision with increasing emitter density demonstrated by the DECODE results and it does so while decreasing the rate at which the detection accuracy decreases with emitter density, further improving temporal resolution.

\begin{figure}[!h]
\centering
\includegraphics[width=0.7\textwidth]{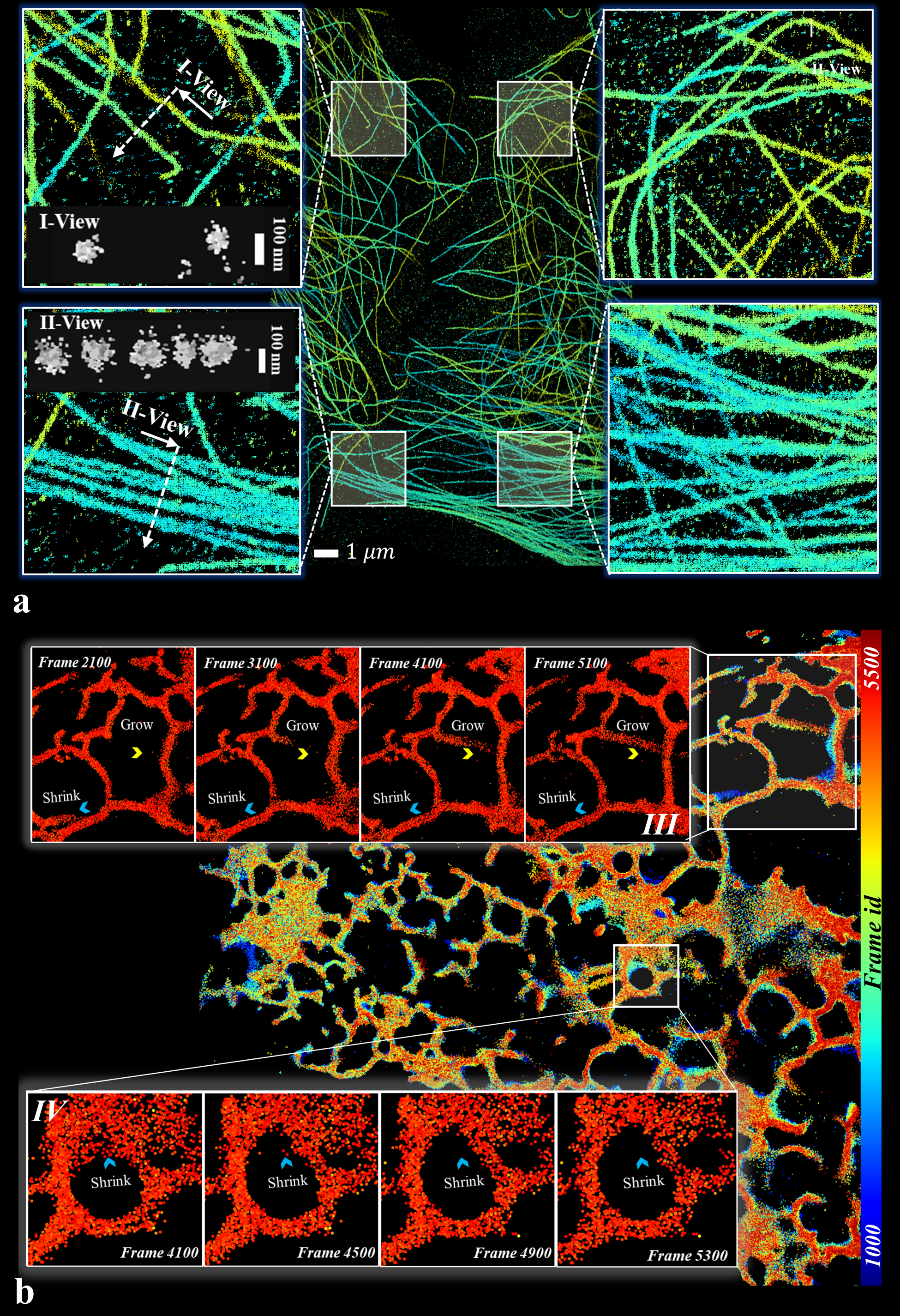}
\caption{\textbf{LUENN Reconstruction performance on ultra-dense labeled Tublins and Fast live-cell.} \textbf{a.} Microtubules labeled with a high concentration of anti-$\mathrm{\alpha}$ and anti-$\mathrm{\beta}$-tubulin primary and Alexa Fluor 647 secondary antibodies. The camera used to record the data has a pixel size of 117 × 127 nm. Side view reconstructions of magnified regions as indicated in \textbf{\emph{I}} and \textbf{\emph{II}} views. \textbf{b.} Fast live-cell SMLM on the Golgi apparatus labeled with $\mathrm{\alpha}$-mannosidase II-mEos3.2. Localized seeds for all frames combined and colored by the corresponding frame id. \textbf{II}. Grows and Shrinks are shown by the combining the last 1500 frames and rendered based on the axial location. \textbf{(Supplementary video 1)} \textbf{IV}. Shrinks and the increases of the hole is shown by the combining the last 1500 frames and rendered based on the axial location. \textbf{(Supplementary video 2)}. }\label{fig:Figure_Combined1}
\end{figure}
\begin{figure}[!h]
\centering
\includegraphics[width=0.7\textwidth]{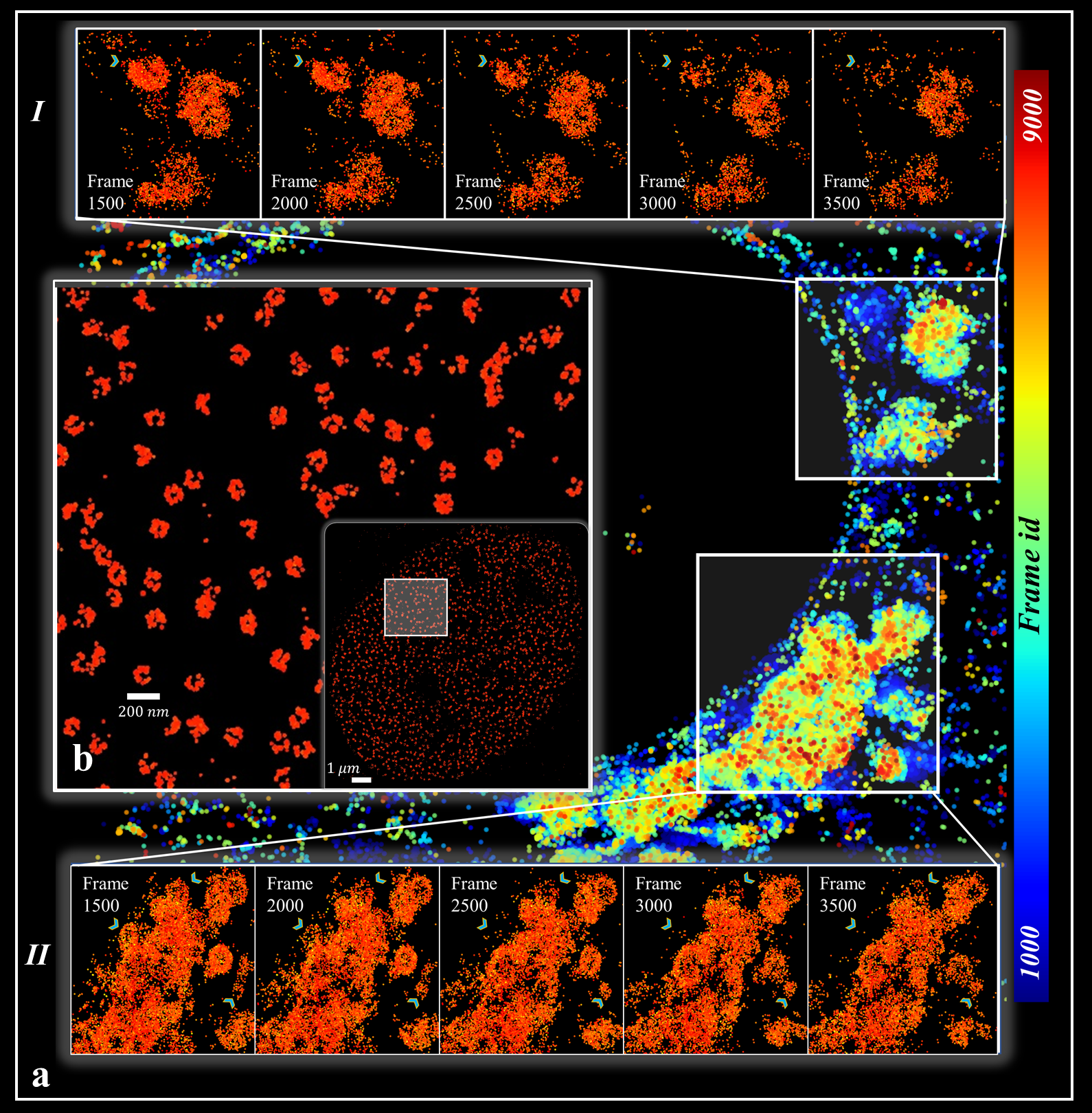}
\caption{\textbf{LUENN Reconstruction performance on Fast live-cell and ultra-high labeled nuclear pores.} \textbf{a.} Fast live-cell SMLM on the endoplasmic reticulum labeled with calnexin-mEos3.2. Localized seeds for all frames combined and colored by the corresponding frame id. \textbf{\emph{I}} and \textbf{\emph{II}} are shown by the combining the last 1500 frames and rendered based on the axial location. \textbf{(Supplementary video 3 and 4)} \textbf{b.} Fast live-cell SMLM on the nuclear pore complex protein Nup96-mMaple acquired in 3s data. Magnified region is shown on the background with 200 nm label size.}\label{fig:Figure_Combined2}
\end{figure}

We have applied LUENN to several datasets provided by SMLM 2016 Challenge, Speiser et al., and Li et al. \citep{sage_super-resolution_2019,decode,li2018real} in Figures \ref{fig:Figure_Combined1} and \ref{fig:Figure_Combined2}. Here the benefits of eliminating pixel-level biases become clear, even in low signal to noise experimental data there is an absence of coercion of results to pixel centers without utilizing additional filtering. Figure \ref{fig:Figure_Combined1}a shows Tublin networks from the SMLM Challenge data set \citep{sage_super-resolution_2019,li2018real} and Figure \ref{fig:Figure_Combined2}b shows reconstructions of nuclear pore complexes (NPCs) \citep{decode}. Figures \ref{fig:Figure_Combined1}a show that the LUENN algorithm
faithfully reconstructs the tublins, including its circular cross-section as shown in the $A-A^{'}$ cross-section. In the NPC reconstruction, the 135 nm pore diameter is also faithfully reconstructed. These results are obtained without filtering, increasing the JI and potentially resulting in improvements to temporal resolution. 

The LUENN algorithm also performs well in live-cell, time resolved localization reconstructions. Figure \ref{fig:Figure_Combined1}b shows fast, live-cell SMLM on the Golgi apparatus labeled with $\mathrm{\alpha}$-mannosidase II-mEos3.2 \citep{decode}. Here one can observe the time evolution of the Golgi apparatus structure in the \textbf{\textit{III}} and \textbf{\textit{IV}} time-series views. Figure \ref{fig:Figure_Combined2}a shows fast, live-cell reconstruction of endoplasmic reticulum labeled with calnexinmEos3.2 \citep{decode}. Here the growth and deconstruction of the endoplasmic reticulum is captured across the experiment duration, \textbf{\textit{I}} and \textbf{\textit{II}} time-series views. The entire time series is presented here on the background, color coded by frame number, clearly showing the evolution of the structure.

\section{Discussion and Conclusion}
The LUENN localization algorithm has a unique, CNN architecture designed to be able to smoothly transition between completely isolated and completely overlapping emitters and which shows no bias. This architecture exposes the uncertainty in the results, permitting the development of a second CNN which provides an accurate estimator of the localization uncertainty. Taken together these result in a 3D localization algorithm that is remarkably robust with respect to emitter density. 

Most striking is LUENN's localization precision behavior with increasing emitter density. All localization algorithms, to our knowledge, suffer from increasingly poor precision as emitter images increasingly overlap. LUENN, in contrast, demonstrates a leveling out of precision at high emitter densities ($\mathrm{> 5 emitters/ \mu m^2}$), permitting a factor of 6 increase in acceptable emitter density over current state-of the art algorithms. When applied to a temporally compressed version of the SMLM Challenge MT0 dataset, this translated into a thirty fold improvement in temporal resolution over the single emitter case with an approximately $\mathrm{30 ~nm}$ penalty to localization precision. 

This improvement in temporal resolution and the penalty to localization precision are robust with respect to emitter density, making experimental design easier. The window of emitter densities that are permissible for an experiment becomes larger, easing the requirements for experiment design and reducing experimental iterations  \citep{hess2006ultra}.
\begin{acknowledgement}

The authors express their gratitude for the assistance and correspondence of Artur Speiser, Dr. Srinivas Turaga, and Dr. Jakob Macke, without their assistance the bench-marking against the current state-of-the-art DECODE algorithm would not have been possible.

\end{acknowledgement}

\begin{suppinfo}

\textbf{LUENN is an auto-encoder network that learns a nonlinear mapping function between low and high-resolution frames.} As is shown in supplementary Figure \ref{fig:model}, LUENN is a very deep convolutional encoder–decoder network with symmetric skip connections. The encoder stage serves as a feature extraction model that represents a compressed knowledge of the input data and extracts independent features of the input frame. We utilized the benefits of existing state-of-the-art VGG19 that specializes in building very deep convolutional networks for large-scale visual recognition. VGG19 contains 16 convolutional layers for learning features and all convolution layers have identical kernel size and stride, 3x3 filter with stride 1, Relu activation function, and one-pixel zero padding. At the end of each convolution blocks, there is a pooling layer with down-sampling factor of 2.\\

The decoder stage in the network that is designed to integrate the information extracted by the encoder stage consists of two steps: the first step takes the output of the encoder stage and up-samples it to the same size as the input frame using four convolution blocks. There are skip connections between the mirrored blocks in this stage and the encoder stage. Each skip connection concatenates the last activation of the encoder block with the inputs of the corresponding block in the decoder. These skip connections are used to pass features from the encoder path to the decoder path in order to recover spatial information lost during down-sampling, enabling feature re-usability and stabilizing gradient updates in deep architectures. This improves the stability of training and the convergence of the model. The second step includes 2 densely connected convolution blocks that each up-samples by a factor of 2 and predict channels of the super-resolution (X4) images in the complex domain. These two channels have the necessary information to localize emitters in three-dimension and evaluate uncertainty of prediction.\\

\begin{figure}[h]
\centering
\includegraphics[width=.9\textwidth]{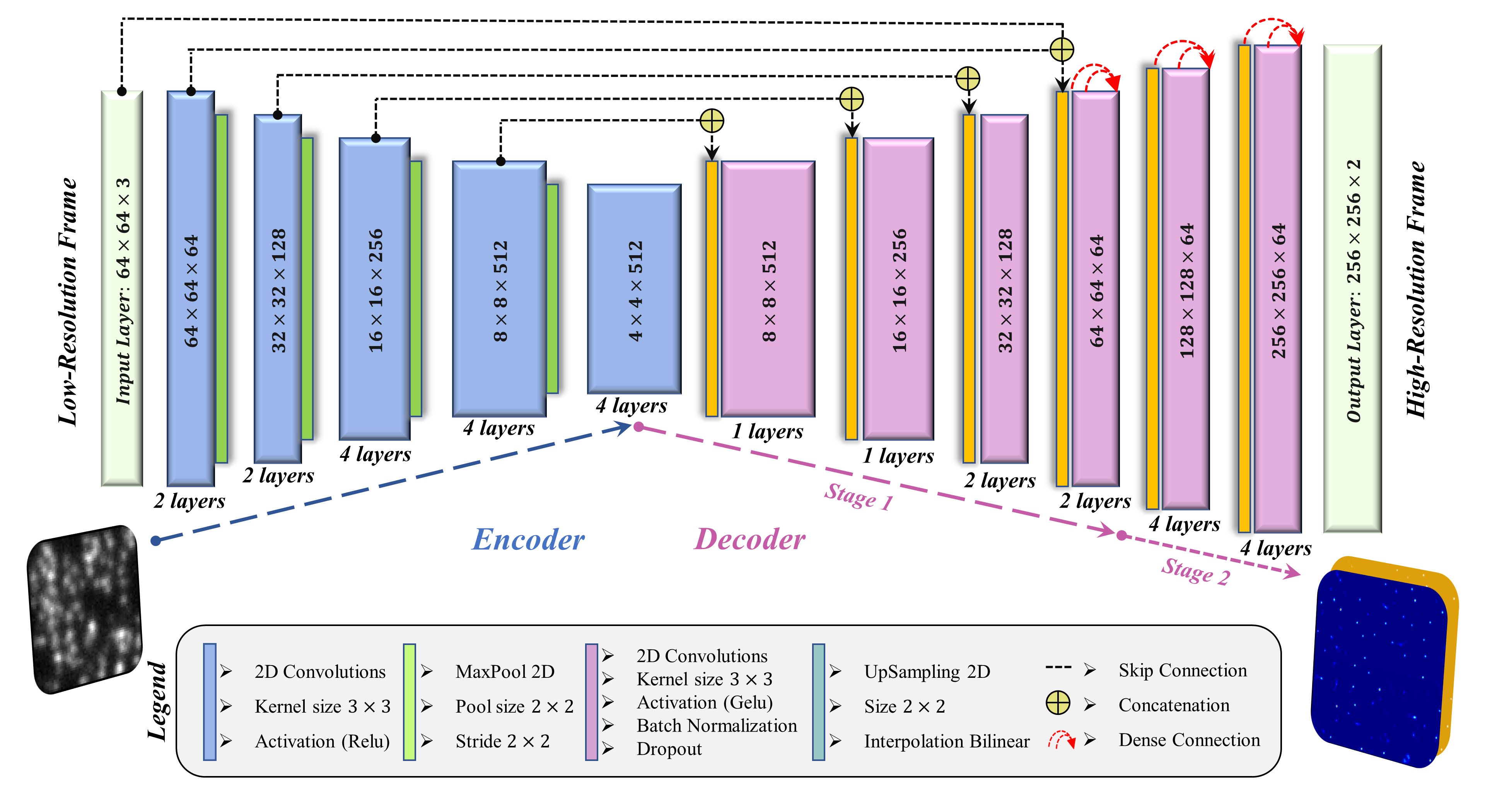}
\caption{\textbf{LUENN architecture.} The proposed LUENN is a deep convolutional encoder-decoder network designed to map low-resolution images to high-resolution images. \textbf{Encoder.} Located on the left side, is VGG-19 model and is responsible for extracting features from the raw input frame. It includes 5 convolution blocks (blue blocks), 16 convolution layers, and 20 million trainable parameters in total. The convolution layers within each block have similar numbers of channels and input sizes, $\mathrm{3x3}$ kernel size with stride 1, and ReLu activation function. Additionally, each block includes a pooling layer (green blocks) that downsamples the image by a factor of 2 and nearest-neighbor interpolation. \textbf{Decoder stage.} Located on the right side, integrates the extracted features and maps them to two high-resolution images in the complex domain. It includes 6 convolution blocks (pink blocks), 14 convolution layers, and 5.7 million trainable parameters. The input layer of each block is an upsampling layer (orange blocks) with a factor of 2, and bi-linear interpolation, that concatenates with the activation of the mirrored block in the encoder stage. The convolution layers consist of a 2D convolution layer with 3x3 kernel size and stride of 1, a GeLU activation function, a Batch Normalization layer, and a Dropout layer. The layers of the last 3 convolution blocks, stage 2, are densely connected for better convergence and robust training.}\label{fig:model}
\end{figure}

\textbf{Step-wise training and loss function.} Image-to-image translation problems are often formulated as pixel-wise regression in which the model learns an end-to-end mapping function between the reconstructed and the labeled images by minimizing a loss function. Mean Squared Error (MSE) is a commonly used loss function for pixel-wise image comparison. As is shown in supplementary Figure \ref{fig:train}a, it calculates the difference between the predicted and the ground truth images by taking the average of the squared differences of each pixel. It is a popular choice for image-to-image translation problems as it provides a clear measure of the difference between the predicted and ground truth images and can be easily optimized using gradient descent methods. For optimization, we employed the ADAM optimization method and scheduled 4 steps, following the transfer learning phase. Except for the first two phases, the learning rate was decreased by half at the start of each phase and kept constant. Supplementary Figure \ref{fig:train}b illustrates the normalized training and validation losses versus the training time. Runtimes were measured on a single NVIDIA Tesla V100 GPU for an example model trained on frames simulated with low emitter density, high SNR, and Astigmatism modality.\\

\begin{figure}[h]
\centering
\includegraphics[width=.7\textwidth]{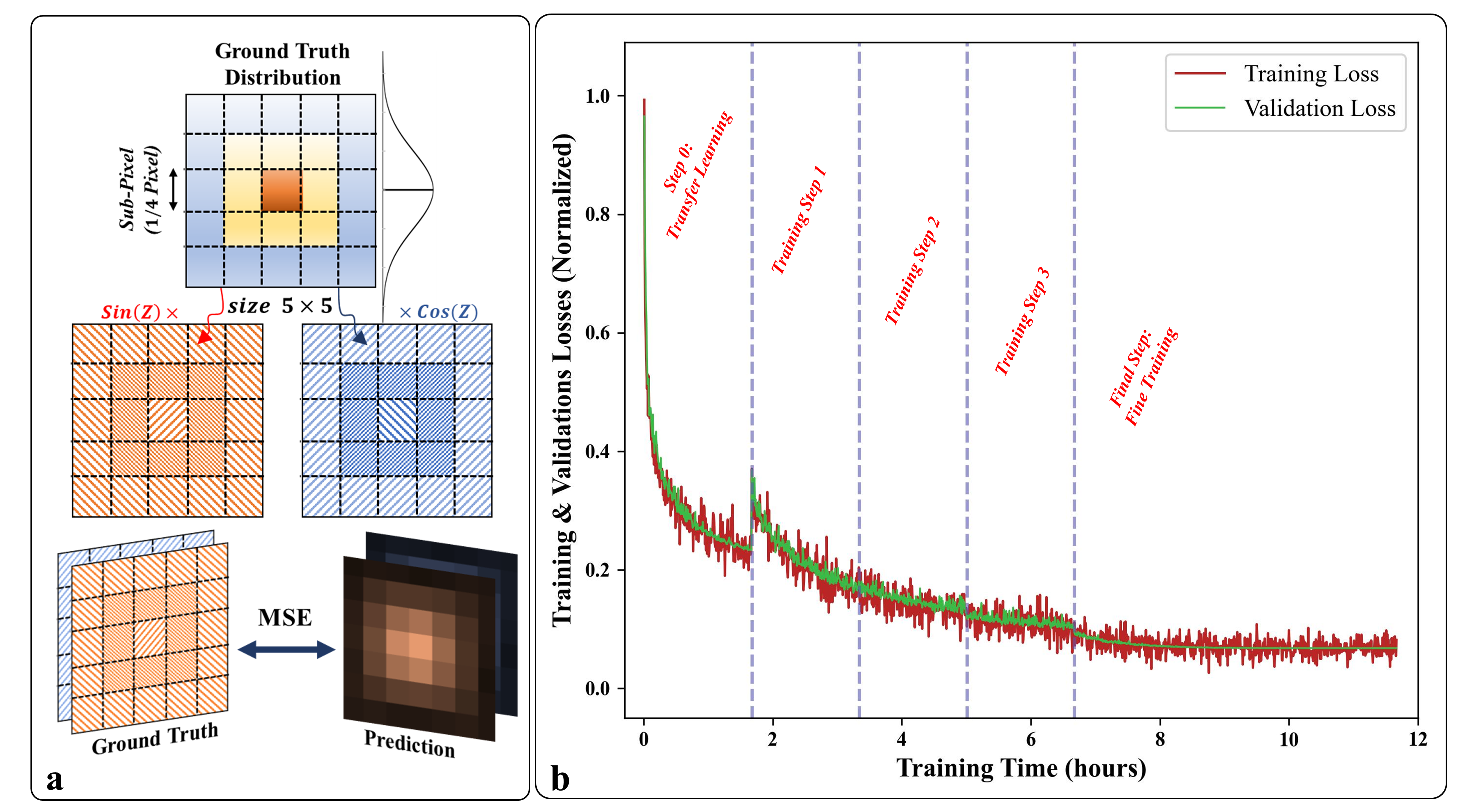}
\caption{\textbf{a. Label example.} Pixel-wise comparison between the Ground Truth distributions in the complex domain and the corresponding model predictions using Mean Squared Error (MSE) loss function. \textbf{b. Step-wise training and losses.} Normalized training and validation losses versus the training time measured on a single NVIDIA Tesla V100 GPU. The results are for an example model trained on frames simulated with low emitter density, high SNR, and Astigmatism modality. 4 steps of training scheduled, following the transfer learning phase, step 0, that each had constant learning-rate equals to half of the previous step.} \label{fig:train}
\end{figure}

\textbf{Training data sampling is challenging.} In deep CNNs, the number of trainable parameters, also known as the parameter domain, is large, which increases the risk of over-fitting as more images with similar features are analyzed per training epoch. Furthermore, in SMLM, the feature domain, which refers to the set of features and their distribution, is vast and diverse, making it challenging to train the model effectively due to the limited number of real experimental frames with ground truth available. To address this challenge, a combination of techniques was employed. First, we used a pre-trained version of VGG-19 on the ImageNet dataset to leverage and transfer its knowledge at the starting phase of the training. Transfer learning helps to improve the stability of training and the generalization of the model on the target domain. Second, we used a training sample generation method during training, using PSF engineering function. On the fly sampling of the training data, ensures that the model does not extract sample-specific features and overfit on specific set of frames.\\

\textbf{Full Computational time for LUENN prediction and localization.} The computational time for a single frame with varying emitter densities was measured on a single NVIDIA RTX 2080 Ti GPU and reported in supplementary Table \ref{tab2:processing_time} both per frame and per seed. The computation time of a frame is summation of the prediction and localization steps. As seen in the table, the processing time per frame increases as the emitter density increases due to the increased number of seeds that need to be localized. However, the processing time per seed gradually decreases. It is important to note that these processing times were measured and averaged over all simulated frames.

\begin{table}[h]
\begin{center}
\begin{minipage}{\textwidth}
\caption{LUENN Localization Processing time}\label{tab2:processing_time}
\begin{tabular*}{\textwidth}{p{0.5in}p{1.5in}p{1.5in}p{1.5in}}

ID & Density ($\mathrm{emitter/\mu m^2}$) & Processing time per seed (ms)& Processing time per frame (ms)\\
\hline \hline
1 & 0.035870  & 16.0 & 11.1 \\
2 & 0.103306  & 8.0 & 16.0 \\
3 & 0.206612  & 7.0 & 27.9 \\
4 & 0.464876  & 4.8 & 43.1 \\
5 & 1.033058  & 2.7 & 53.7 \\
6 & 2.376033  & 1.2 & 57.3 \\
7 & 5.630165  & 0.6 & 66.9 \\
8 & 9.297520  & 0.3 & 54.8 \\
9 & 15.49587  & 0.2 & 50.8 \\
10 & 30.99174  & 0.9 & 85.9 \\

\end{tabular*}
\end{minipage}
\end{center}
\end{table}

\textbf{How much data is enough?} It is an important question in any data-based statistical approach to determine how much data is "enough." The amount of data affects the computational time and the conclusions of the analysis. It is required to have confidence in our assessment without wasting resources. Insufficient data may lead to inaccurate conclusions and evaluations, whereas excessive data leads to computational complexity. In SMLM, frames have varying numbers of emitters, or emitter densities, which each contribute as an example in the localization assessment. In evaluation, the precision of localization approaches is measured by volumetric (3D) and lateral (2D) RMSE, which are averaged over the detected emitters. Statistically speaking, RMSE changes dramatically if the number of outliers increases and its value is reliable when the number of data samples covers the entire range of possible values. However, this question has never been answered in previous works, and authors have simply reported the number of simulated frames and it is not clear why those example frames are enough for their evaluations. If more frames were simulated, would the localization precision change or not? To answer this question from a statistical perspective, we first simulated 100 frames and measured localization precision parameters. Then, more frames were added one by one to the pile and processed. In this way, we could plot the residuals of the parameters by ongoing simulations and stop when the residuals were negligible. In supplementary Figure \ref{fig:residuals}, we plotted on-going calculations of Jaccardian Index, Lateral, and Volumetric RMSE versus the detected emitters. With this approach, we found that at least 25,000 seeds are required and it is independent of the frame density. 

\begin{figure}[!h]
\centering
\includegraphics[width=0.7\textwidth]{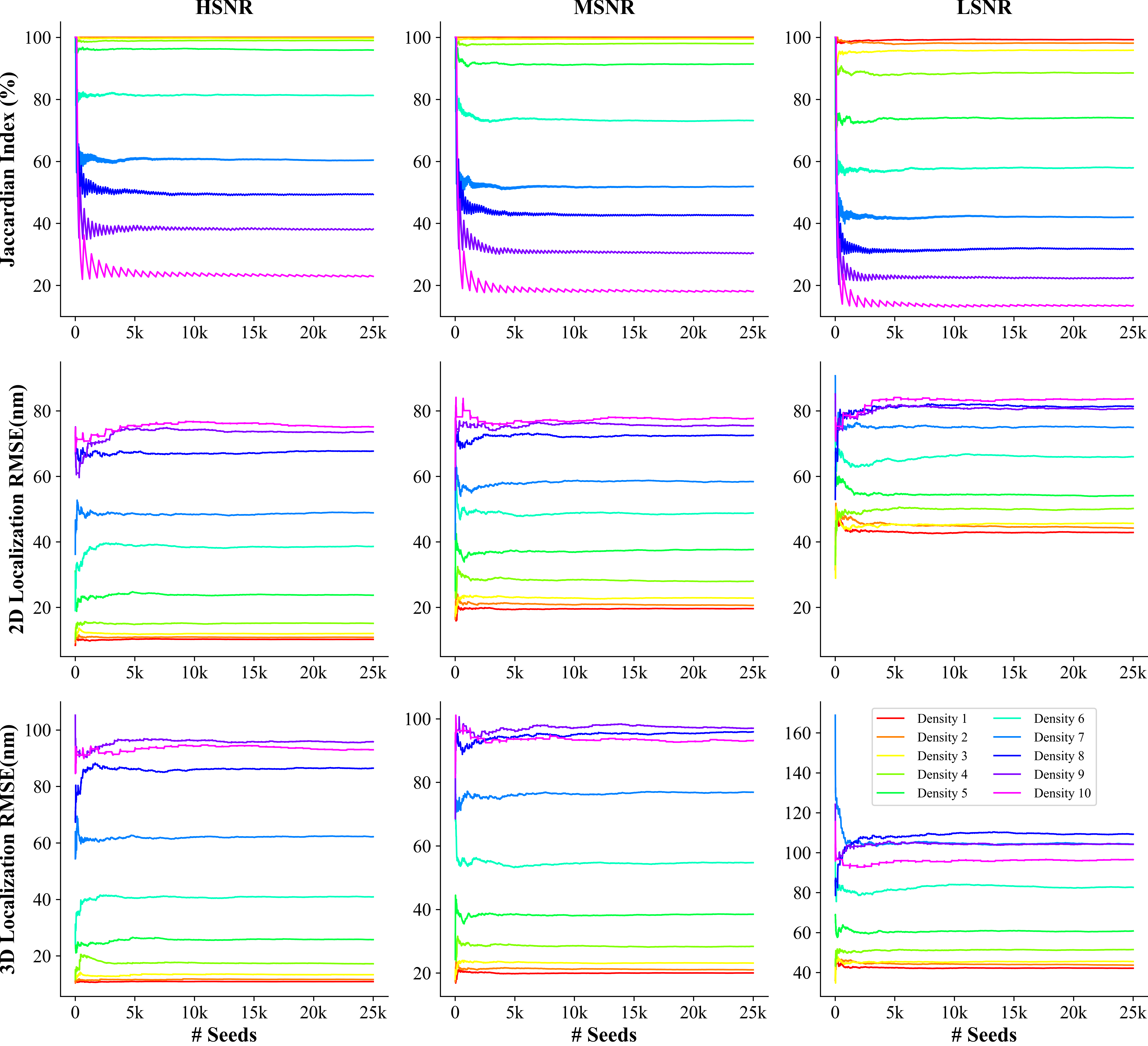}
\caption{\textbf{Residual plots.} Statistical evaluation parameters, Jaccardian Index, Volumetric RMSE, and Lateral RMSE, are plotted versus the number of detected seeds during simulation-evaluation for the three SNR levels, Low, Medium, and High SNRs.}\label{fig:residuals}
\end{figure}
\textbf{Uncertainty estimation and the impact of filtering rates on localization error and detection accuracy.} The impact of filtering on localization error, volumetric RMSE, and detection accuracy, Jaccardian Index, are demonstrated in supplementary Figure \ref{fig:filters}. Each line is localization results for specific density and varying of filtering rates, ranging from 0 to 85 percent. In any rate, a percentage of worst predictions that are estimated by the uncertainty model are removed. It can be observed that as the filtering rate increases, the localization precision improves while the detection accuracy decreases. This means that by removing a higher percentage of predictions with higher estimated precision or uncertainty, the overall localization accuracy of the remaining predictions increases, but the number of detections decreases. The optimal filtering rate will depend on the specific application and the desired trade-off between localization precision and detection efficiency.

\begin{figure}[!h]
\centering
\includegraphics[width=0.7\textwidth]{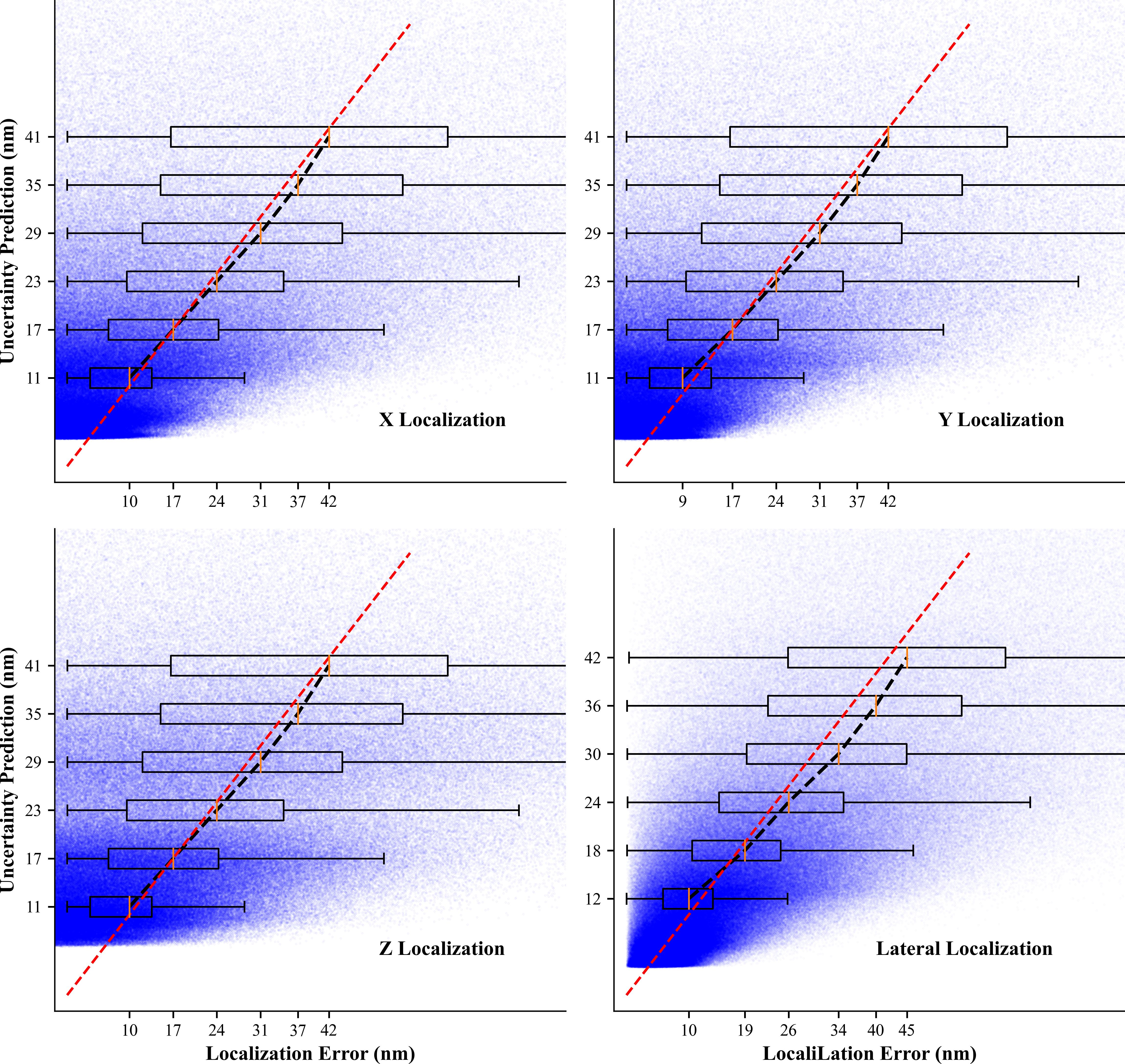}
\caption{\textbf{Localization Error estimations} LUENN uncertainty prediction model is able to estimate localization error in X, Y, and Z directions separately that could be used for lateral and total error estimations. In these plots, the blue points are the detected seeds from frames with low, Medium, and High SNRs and densities of 1 and 2.4.}\label{fig:unc}
\end{figure}

\begin{figure}[!h]
\centering
\includegraphics[width=0.7\textwidth]{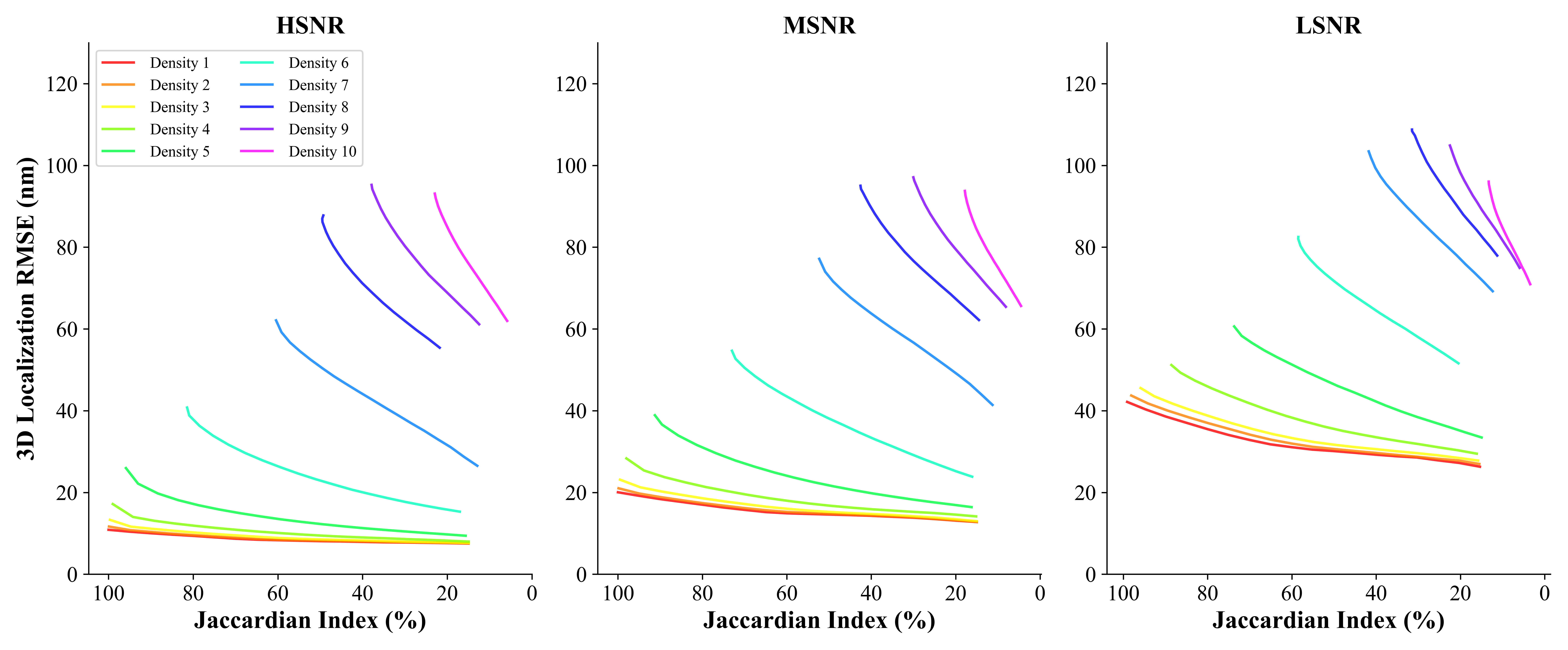}
\caption{\textbf{LUENN Uncertainty estimation and the impact of filtering rates on localization error and detection accuracy.} Statistical evaluation parameters, Jaccardian Index, Volumetric RMSE, and Lateral RMSE, are plotted versus the number of detected seeds during simulation-evaluation for the three SNR levels, Low, Medium, and High SNRs.}\label{fig:filters}
\end{figure}

\begin{table}[!h]
\begin{center}
\begin{minipage}{\textwidth}
\caption{Different LUENN versions trained for the provided  results.}\label{tab2:models}
\begin{tabular*}{\textwidth}{p{0.8in}p{0.9in}p{.9in}p{0.9in}p{0.5in}p{.6in}}

Model &  Intensity ($\mu,\sigma$) & Training Density ($1/\mu^2$) & Evaluation Density ($1/\mu^2$)& SNR & Ref. Figure \\
\hline \hline
AI-1 &
1k,50 & 0.77 & 0.04-1.0 & Low & \ref{fig:results} \& \ref{fig:decode_compare} \\
AI-2 &
 5k,250 & 0.77 & 0.04-1.0 & Medium & \ref{fig:results} \& \ref{fig:decode_compare} \\
AI-3 &
 20k,1k & 0.77 & 0.04-1.0 & High & \ref{fig:results} \& \ref{fig:decode_compare} \\
AI-4 &
 1k,50 & 4.13 & 2.4-5.47 & Low & \ref{fig:results} \& \ref{fig:decode_compare} \\
AI-5 &
 5k,250 & 4.13 & 2.4-5.47 & Medium & \ref{fig:results} \& \ref{fig:decode_compare} \\
AI-6 &
 20k,1k & 4.13 & 2.4-5.47 & High & \ref{fig:results} \& \ref{fig:decode_compare} \\
AI-7 &
1k,50 & 15.5 & 9.3-31 & Low & \ref{fig:results} \& \ref{fig:decode_compare} \\
AI-8 &
 5k,250 & 15.5 & 9.3-31 & Medium & \ref{fig:results} \& \ref{fig:decode_compare} \\
AI-9 &
 20k,1k & 15.5 & 9.3-31 & High & \ref{fig:results} \& \ref{fig:decode_compare} \\
AI-Tublins &
 7k,1k & 1.22 & NA & Medium & \ref{fig:Figure_Combined1}a\\
 AI-IiveCell &
 5k,1.5k & 0.85 & NA & Medium & \ref{fig:Figure_Combined1}b \& \ref{fig:Figure_Combined2}a\\
AI-NPC &
2.5k,700 & 1.22 & NA & Medium & \ref{fig:Figure_Combined2}b\\
AI-AS &
1k,300 & 0.62 & 0.31 & Low & Sup. \ref{fig:challenge}a\\
AI-DH &
3.6k,1k & 0.62 & 0.31 & Low & Sup. \ref{fig:challenge}b\\

\end{tabular*}
\end{minipage}
\end{center}
\end{table}

\textbf{Reconstruction quality of SMLM 2016 datasets.} The performance of LUENN on SMLM 2016 challenge datasets are shown in supplementary Figure\ref{fig:challenge}. These datasets were simulated with high emitter density and low SNR in astigmatism (left) and double helix (right) modalities. The results demonstrate that LUENN is able to accurately reconstruct the ground truth images and achieve high localization precision in both modalities. The high emitter density and low SNR in these datasets are the most challenging conditions for localization-based methods, making the results of LUENN even more impressive. Overall, the results show that LUENN is a robust and efficient algorithm for super-resolution microscopy, capable of handling a wide range of imaging conditions.
\begin{figure}[!h]
\centering
\includegraphics[width=0.6\textwidth]{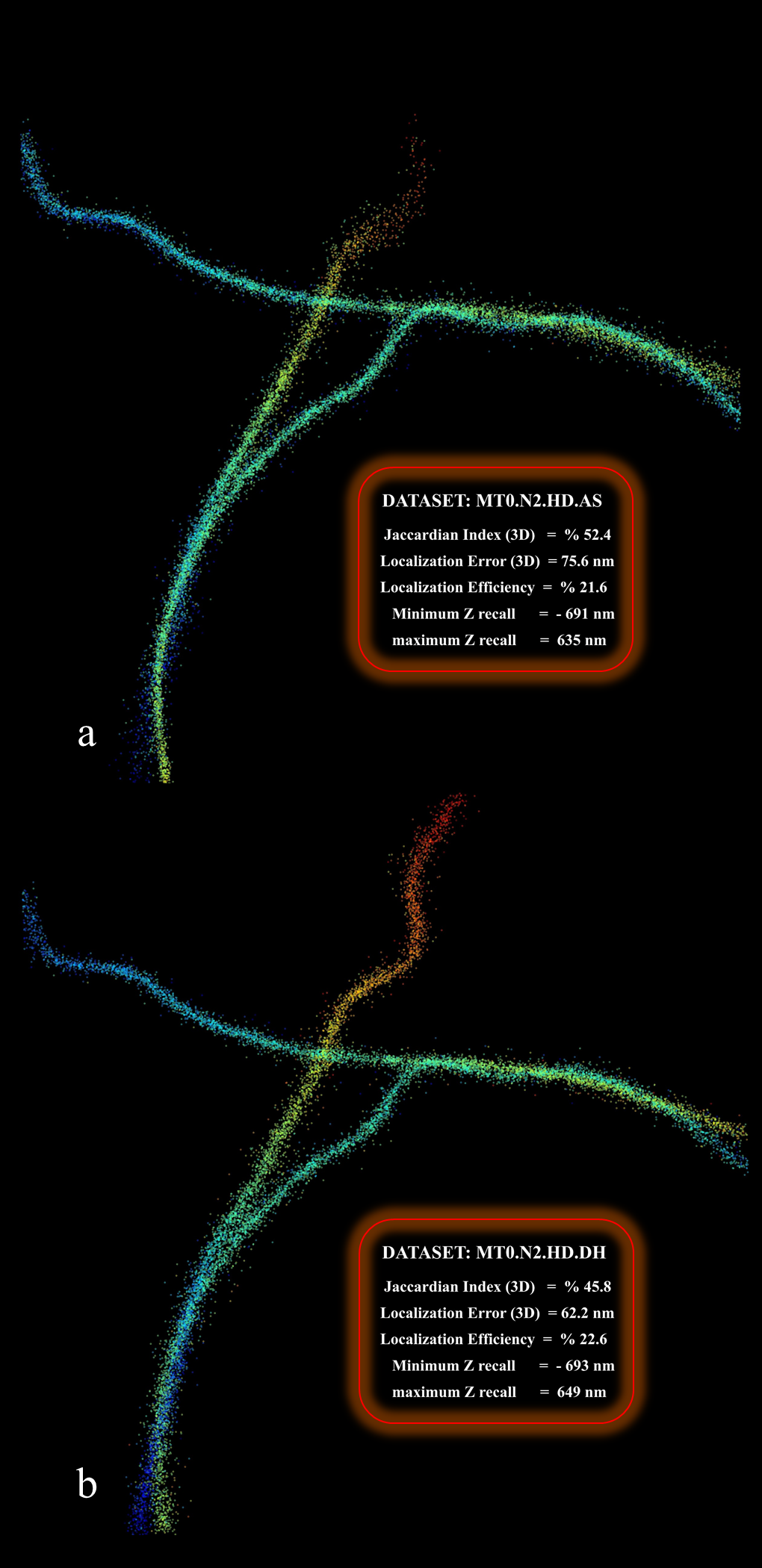}
\caption{LUENN localization and reconstruction performance on SMLM 2016 Challenge datasets in (a) Astigmatism and (b) Double-Helix modalities in the most challenging condition of low SNR and high emitter density.}\label{fig:challenge}
\end{figure}

\end{suppinfo}

\bibliography{achemso-demo}

\providecommand{\latin}[1]{#1}
\makeatletter
\providecommand{\doi}
  {\begingroup\let\do\@makeother\dospecials
  \catcode`\{=1 \catcode`\}=2 \doi@aux}
\providecommand{\doi@aux}[1]{\endgroup\texttt{#1}}
\makeatother
\providecommand*\mcitethebibliography{\thebibliography}
\csname @ifundefined\endcsname{endmcitethebibliography}
  {\let\endmcitethebibliography\endthebibliography}{}
\begin{mcitethebibliography}{27}
\providecommand*\natexlab[1]{#1}
\providecommand*\mciteSetBstSublistMode[1]{}
\providecommand*\mciteSetBstMaxWidthForm[2]{}
\providecommand*\mciteBstWouldAddEndPuncttrue
  {\def\EndOfBibitem{\unskip.}}
\providecommand*\mciteBstWouldAddEndPunctfalse
  {\let\EndOfBibitem\relax}
\providecommand*\mciteSetBstMidEndSepPunct[3]{}
\providecommand*\mciteSetBstSublistLabelBeginEnd[3]{}
\providecommand*\EndOfBibitem{}
\mciteSetBstSublistMode{f}
\mciteSetBstMaxWidthForm{subitem}{(\alph{mcitesubitemcount})}
\mciteSetBstSublistLabelBeginEnd
  {\mcitemaxwidthsubitemform\space}
  {\relax}
  {\relax}

\bibitem[Rust \latin{et~al.}(2006)Rust, Bates, and Zhuang]{storm}
Rust,~M.~J.; Bates,~M.; Zhuang,~X. Sub-Diffraction-Limit Imaging by Stochastic
  Optical Reconstruction Microscopy ({{STORM}}). \emph{Nature Methods}
  \textbf{2006}, \emph{3}, 793--796\relax
\mciteBstWouldAddEndPuncttrue
\mciteSetBstMidEndSepPunct{\mcitedefaultmidpunct}
{\mcitedefaultendpunct}{\mcitedefaultseppunct}\relax
\EndOfBibitem
\bibitem[Betzig \latin{et~al.}(2006)Betzig, Patterson, Sougrat, Lindwasser,
  Olenych, Bonifacino, Davidson, {Lippincott-Schwartz}, and Hess]{palm}
Betzig,~E.; Patterson,~G.~H.; Sougrat,~R.; Lindwasser,~O.~W.; Olenych,~S.;
  Bonifacino,~J.~S.; Davidson,~M.~W.; {Lippincott-Schwartz},~J.; Hess,~H.~F.
  Imaging Intracellular Fluorescent Proteins at Nanometer Resolution.
  \emph{Science (New York, N.Y.)} \textbf{2006}, \emph{313}, 1642--5\relax
\mciteBstWouldAddEndPuncttrue
\mciteSetBstMidEndSepPunct{\mcitedefaultmidpunct}
{\mcitedefaultendpunct}{\mcitedefaultseppunct}\relax
\EndOfBibitem
\bibitem[Hell and Wichmann(1994)Hell, and Wichmann]{sted}
Hell,~S.~W.; Wichmann,~J. Breaking the Diffraction Resolution Limit by
  Stimulated Emission: Stimulated-Emission-Depletion Fluorescence Microscopy.
  \emph{Optics letters} \textbf{1994}, \emph{19}, 780--2\relax
\mciteBstWouldAddEndPuncttrue
\mciteSetBstMidEndSepPunct{\mcitedefaultmidpunct}
{\mcitedefaultendpunct}{\mcitedefaultseppunct}\relax
\EndOfBibitem
\bibitem[Gustafsson(2000)]{gustafsson_surpassing_2000}
Gustafsson,~M. G.~L. Surpassing the Lateral Resolution Limit by a Factor of Two
  Using Structured Illumination Microscopy. {{SHORT COMMUNICATION}}.
  \emph{Journal of Microscopy} \textbf{2000}, \emph{198}, 82--87\relax
\mciteBstWouldAddEndPuncttrue
\mciteSetBstMidEndSepPunct{\mcitedefaultmidpunct}
{\mcitedefaultendpunct}{\mcitedefaultseppunct}\relax
\EndOfBibitem
\bibitem[Hell and Stelzer(1992)Hell, and Stelzer]{hell_properties_1992}
Hell,~S.; Stelzer,~E. H.~K. Properties of a {{4Pi}} Confocal Fluorescence
  Microscope. \emph{JOSA A} \textbf{1992}, \emph{9}, 2159--2166\relax
\mciteBstWouldAddEndPuncttrue
\mciteSetBstMidEndSepPunct{\mcitedefaultmidpunct}
{\mcitedefaultendpunct}{\mcitedefaultseppunct}\relax
\EndOfBibitem
\bibitem[Sharonov and Hochstrasser(2006)Sharonov, and
  Hochstrasser]{sharonov_wide-field_2006}
Sharonov,~A.; Hochstrasser,~R.~M. Wide-Field Subdiffraction Imaging by
  Accumulated Binding of Diffusing Probes. \emph{Proceedings of the National
  Academy of Sciences of the United States of America} \textbf{2006},
  \emph{103}, 18911--18916\relax
\mciteBstWouldAddEndPuncttrue
\mciteSetBstMidEndSepPunct{\mcitedefaultmidpunct}
{\mcitedefaultendpunct}{\mcitedefaultseppunct}\relax
\EndOfBibitem
\bibitem[Liu \latin{et~al.}(2022)Liu, Hoess, and
  Ries]{liu_super-resolution_2022}
Liu,~S.; Hoess,~P.; Ries,~J. Super-{{Resolution Microscopy}} for {{Structural
  Cell Biology}}. \emph{Annual Review of Biophysics} \textbf{2022}, \emph{51},
  301--326\relax
\mciteBstWouldAddEndPuncttrue
\mciteSetBstMidEndSepPunct{\mcitedefaultmidpunct}
{\mcitedefaultendpunct}{\mcitedefaultseppunct}\relax
\EndOfBibitem
\bibitem[Lelek \latin{et~al.}(2021)Lelek, Gyparaki, Beliu, Schueder,
  Griffi{\'e}, Manley, Jungmann, Sauer, Lakadamyali, and
  Zimmer]{lelek_single-molecule_2021}
Lelek,~M.; Gyparaki,~M.~T.; Beliu,~G.; Schueder,~F.; Griffi{\'e},~J.;
  Manley,~S.; Jungmann,~R.; Sauer,~M.; Lakadamyali,~M.; Zimmer,~C.
  Single-Molecule Localization Microscopy. \emph{Nature Reviews Methods
  Primers} \textbf{2021}, \emph{1}, 1--27\relax
\mciteBstWouldAddEndPuncttrue
\mciteSetBstMidEndSepPunct{\mcitedefaultmidpunct}
{\mcitedefaultendpunct}{\mcitedefaultseppunct}\relax
\EndOfBibitem
\bibitem[Sage \latin{et~al.}(2019)Sage, Pham, Babcock, Lukes, Pengo, Chao,
  Velmurugan, Herbert, Agrawal, Colabrese, Wheeler, Archetti, Rieger, Ober,
  Hagen, Sibarita, Ries, Henriques, Unser, and
  Holden]{sage_super-resolution_2019}
Sage,~D. \latin{et~al.}  Super-Resolution Fight Club: Assessment of {{2D}} and
  {{3D}} Single-Molecule Localization Microscopy Software. \emph{Nature
  Methods} \textbf{2019}, \emph{16}, 387--395\relax
\mciteBstWouldAddEndPuncttrue
\mciteSetBstMidEndSepPunct{\mcitedefaultmidpunct}
{\mcitedefaultendpunct}{\mcitedefaultseppunct}\relax
\EndOfBibitem
\bibitem[Liu \latin{et~al.}(2018)Liu, Qiao, Wu, Wang, Fang, and
  Dai]{liu2018fast}
Liu,~K.; Qiao,~H.; Wu,~J.; Wang,~H.; Fang,~L.; Dai,~Q. Fast 3D cell tracking
  with wide-field fluorescence microscopy through deep learning. \emph{arXiv
  preprint arXiv:1805.05139} \textbf{2018}, \relax
\mciteBstWouldAddEndPunctfalse
\mciteSetBstMidEndSepPunct{\mcitedefaultmidpunct}
{}{\mcitedefaultseppunct}\relax
\EndOfBibitem
\bibitem[Boyd \latin{et~al.}(2018)Boyd, Jonas, Babcock, and
  Recht]{boyd2018deeploco}
Boyd,~N.; Jonas,~E.; Babcock,~H.; Recht,~B. DeepLoco: fast 3D localization
  microscopy using neural networks. \emph{BioRxiv} \textbf{2018}, 267096\relax
\mciteBstWouldAddEndPuncttrue
\mciteSetBstMidEndSepPunct{\mcitedefaultmidpunct}
{\mcitedefaultendpunct}{\mcitedefaultseppunct}\relax
\EndOfBibitem
\bibitem[Nehme \latin{et~al.}(2020)Nehme, Freedman, Gordon, Ferdman, Weiss,
  Alalouf, Naor, Orange, Michaeli, and Shechtman]{nehme2020deepstorm3d}
Nehme,~E.; Freedman,~D.; Gordon,~R.; Ferdman,~B.; Weiss,~L.~E.; Alalouf,~O.;
  Naor,~T.; Orange,~R.; Michaeli,~T.; Shechtman,~Y. DeepSTORM3D: dense 3D
  localization microscopy and PSF design by deep learning. \emph{Nature
  methods} \textbf{2020}, \emph{17}, 734--740\relax
\mciteBstWouldAddEndPuncttrue
\mciteSetBstMidEndSepPunct{\mcitedefaultmidpunct}
{\mcitedefaultendpunct}{\mcitedefaultseppunct}\relax
\EndOfBibitem
\bibitem[Speiser \latin{et~al.}(2021)Speiser, M{\"u}ller, Hoess, Matti, Obara,
  Legant, Kreshuk, Macke, Ries, and Turaga]{decode}
Speiser,~A.; M{\"u}ller,~L.-R.; Hoess,~P.; Matti,~U.; Obara,~C.~J.;
  Legant,~W.~R.; Kreshuk,~A.; Macke,~J.~H.; Ries,~J.; Turaga,~S.~C. Deep
  Learning Enables Fast and Dense Single-Molecule Localization with High
  Accuracy. \emph{Nature Methods} \textbf{2021}, \emph{18}, 1082--1090\relax
\mciteBstWouldAddEndPuncttrue
\mciteSetBstMidEndSepPunct{\mcitedefaultmidpunct}
{\mcitedefaultendpunct}{\mcitedefaultseppunct}\relax
\EndOfBibitem
\bibitem[Sugawara \latin{et~al.}(2019)Sugawara, Shiota, and Kiya]{artifacts}
Sugawara,~Y.; Shiota,~S.; Kiya,~H. Checkerboard Artifacts Free Convolutional
  Neural Networks. \emph{APSIPA Transactions on Signal and Information
  Processing} \textbf{2019}, \emph{8}, e9\relax
\mciteBstWouldAddEndPuncttrue
\mciteSetBstMidEndSepPunct{\mcitedefaultmidpunct}
{\mcitedefaultendpunct}{\mcitedefaultseppunct}\relax
\EndOfBibitem
\bibitem[Odena \latin{et~al.}(2016)Odena, Dumoulin, and Olah]{checkerboard}
Odena,~A.; Dumoulin,~V.; Olah,~C. Deconvolution and {{Checkerboard Artifacts}}.
  \emph{Distill} \textbf{2016}, \emph{1}, e3\relax
\mciteBstWouldAddEndPuncttrue
\mciteSetBstMidEndSepPunct{\mcitedefaultmidpunct}
{\mcitedefaultendpunct}{\mcitedefaultseppunct}\relax
\EndOfBibitem
\bibitem[Kinoshita and Kiya(2020)Kinoshita, and Kiya]{Fixed}
Kinoshita,~Y.; Kiya,~H. Fixed smooth convolutional layer for avoiding
  checkerboard artifacts in CNNs. ICASSP 2020-2020 IEEE International
  Conference on Acoustics, Speech and Signal Processing (ICASSP). 2020; pp
  3712--3716\relax
\mciteBstWouldAddEndPuncttrue
\mciteSetBstMidEndSepPunct{\mcitedefaultmidpunct}
{\mcitedefaultendpunct}{\mcitedefaultseppunct}\relax
\EndOfBibitem
\bibitem[Li \latin{et~al.}(2020)Li, Huang, Pei, Jiao, and Shang]{interpolation}
Li,~Y.; Huang,~Q.; Pei,~X.; Jiao,~L.; Shang,~R. RADet: Refine feature pyramid
  network and multi-layer attention network for arbitrary-oriented object
  detection of remote sensing images. \emph{Remote Sensing} \textbf{2020},
  \emph{12}, 389\relax
\mciteBstWouldAddEndPuncttrue
\mciteSetBstMidEndSepPunct{\mcitedefaultmidpunct}
{\mcitedefaultendpunct}{\mcitedefaultseppunct}\relax
\EndOfBibitem
\bibitem[Aitken \latin{et~al.}(2017)Aitken, Ledig, Theis, Caballero, Wang, and
  Shi]{subconv}
Aitken,~A.; Ledig,~C.; Theis,~L.; Caballero,~J.; Wang,~Z.; Shi,~W. Checkerboard
  artifact free sub-pixel convolution: A note on sub-pixel convolution, resize
  convolution and convolution resize. \emph{arXiv preprint arXiv:1707.02937}
  \textbf{2017}, \relax
\mciteBstWouldAddEndPunctfalse
\mciteSetBstMidEndSepPunct{\mcitedefaultmidpunct}
{}{\mcitedefaultseppunct}\relax
\EndOfBibitem
\bibitem[Isola \latin{et~al.}(2017)Isola, Zhu, Zhou, and Efros]{Unet}
Isola,~P.; Zhu,~J.-Y.; Zhou,~T.; Efros,~A.~A. Image-to-{{Image Translation}}
  with {{Conditional Adversarial Networks}}. 2017 {{IEEE Conference}} on
  {{Computer Vision}} and {{Pattern Recognition}} ({{CVPR}}). 2017; pp
  5967--5976\relax
\mciteBstWouldAddEndPuncttrue
\mciteSetBstMidEndSepPunct{\mcitedefaultmidpunct}
{\mcitedefaultendpunct}{\mcitedefaultseppunct}\relax
\EndOfBibitem
\bibitem[M{\"o}ckl \latin{et~al.}(2020)M{\"o}ckl, Roy, Petrov, and
  Moerner]{mockl_accurate_2020}
M{\"o}ckl,~L.; Roy,~A.~R.; Petrov,~P.~N.; Moerner,~W.~E. Accurate and Rapid
  Background Estimation in Single-Molecule Localization Microscopy Using the
  Deep Neural Network {{BGnet}}. \emph{Proceedings of the National Academy of
  Sciences} \textbf{2020}, \emph{117}, 60--67\relax
\mciteBstWouldAddEndPuncttrue
\mciteSetBstMidEndSepPunct{\mcitedefaultmidpunct}
{\mcitedefaultendpunct}{\mcitedefaultseppunct}\relax
\EndOfBibitem
\bibitem[Hyun and Kim(2022)Hyun, and Kim]{hyun_development_2022}
Hyun,~Y.; Kim,~D. Development of {{Deep-Learning-Based Single-Molecule
  Localization Image Analysis}}. \emph{International Journal of Molecular
  Sciences} \textbf{2022}, \emph{23}, 6896\relax
\mciteBstWouldAddEndPuncttrue
\mciteSetBstMidEndSepPunct{\mcitedefaultmidpunct}
{\mcitedefaultendpunct}{\mcitedefaultseppunct}\relax
\EndOfBibitem
\bibitem[Zelger \latin{et~al.}(2018)Zelger, Kaser, Rossboth, Velas, Sch{\"u}tz,
  and Jesacher]{zelger_three-dimensional_2018-1}
Zelger,~P.; Kaser,~K.; Rossboth,~B.; Velas,~L.; Sch{\"u}tz,~G.~J.; Jesacher,~A.
  Three-Dimensional Localization Microscopy Using Deep Learning. \emph{Optics
  Express} \textbf{2018}, \emph{26}, 33166--33179\relax
\mciteBstWouldAddEndPuncttrue
\mciteSetBstMidEndSepPunct{\mcitedefaultmidpunct}
{\mcitedefaultendpunct}{\mcitedefaultseppunct}\relax
\EndOfBibitem
\bibitem[Nikolova \latin{et~al.}(2011)Nikolova, Iliev, Ovtcharov, and
  Poulkov]{digital}
Nikolova,~Z.; Iliev,~G.; Ovtcharov,~M.; Poulkov,~V. In \emph{Applications of
  {{Digital Signal Processing}}}; {Cuadrado-Laborde},~C., Ed.; {InTech},
  2011\relax
\mciteBstWouldAddEndPuncttrue
\mciteSetBstMidEndSepPunct{\mcitedefaultmidpunct}
{\mcitedefaultendpunct}{\mcitedefaultseppunct}\relax
\EndOfBibitem
\bibitem[Li \latin{et~al.}(1999)Li, Valentine, and Rana]{li1999modified}
Li,~J.; Valentine,~J.~D.; Rana,~A.~E. The modified three point Gaussian method
  for determining Gaussian peak parameters. \emph{Nuclear Instruments and
  Methods in Physics Research Section A: Accelerators, Spectrometers, Detectors
  and Associated Equipment} \textbf{1999}, \emph{422}, 438--443\relax
\mciteBstWouldAddEndPuncttrue
\mciteSetBstMidEndSepPunct{\mcitedefaultmidpunct}
{\mcitedefaultendpunct}{\mcitedefaultseppunct}\relax
\EndOfBibitem
\bibitem[Li \latin{et~al.}(2018)Li, Mund, Hoess, Deschamps, Matti, Nijmeijer,
  Sabinina, Ellenberg, Schoen, and Ries]{li2018real}
Li,~Y.; Mund,~M.; Hoess,~P.; Deschamps,~J.; Matti,~U.; Nijmeijer,~B.;
  Sabinina,~V.~J.; Ellenberg,~J.; Schoen,~I.; Ries,~J. Real-time 3D
  single-molecule localization using experimental point spread functions.
  \emph{Nature methods} \textbf{2018}, \emph{15}, 367--369\relax
\mciteBstWouldAddEndPuncttrue
\mciteSetBstMidEndSepPunct{\mcitedefaultmidpunct}
{\mcitedefaultendpunct}{\mcitedefaultseppunct}\relax
\EndOfBibitem
\bibitem[Hess \latin{et~al.}(2006)Hess, Girirajan, and Mason]{hess2006ultra}
Hess,~S.~T.; Girirajan,~T. P.~K.; Mason,~M.~D. Ultra-{{High Resolution
  Imaging}} by {{Fluorescence Photoactivation Localization Microscopy}}.
  \emph{Biophysical Journal} \textbf{2006}, \emph{91}, 4258\relax
\mciteBstWouldAddEndPuncttrue
\mciteSetBstMidEndSepPunct{\mcitedefaultmidpunct}
{\mcitedefaultendpunct}{\mcitedefaultseppunct}\relax
\EndOfBibitem
\end{mcitethebibliography}

\end{document}